\documentclass[12pt]{iopart}
\usepackage{iopams}
\usepackage{bm}
\usepackage{subfigure}
\usepackage{booktabs}
\usepackage{graphicx}
\usepackage{setstack}
\usepackage{cite}
\usepackage[colorlinks, citecolor = blue, urlcolor = blue]{hyperref}

\def\ben{\begin{equation}}
\def\een{\end{equation}}

\newcommand{\mc}[1]{\mathcal{#1}}
\newcommand{\mb}[1]{\mathbf{#1}}

\newcommand{\ui}{\rmi}

\begin{document}

\title[Second response theory]{Second response theory: A theoretical formalism for the propagation of quantum superpositions}

\author{Mart\'in A. Mosquera}
\address{Department of Chemistry and Biochemistry, Montana State University, Bozeman, MT 59717, USA}
\ead{martinmosquera@montana.edu}

\begin{abstract}  
The propagation of general electronic quantum states provides information of the interaction of
molecular systems with external driving fields. These can also offer understandings regarding
non-adiabatic quantum phenomena. Well established methods focus mainly on propagating a quantum
system that is initially described exclusively by the ground state wavefunction. In this work, we
expand a previously developed size-extensive formalism within coupled cluster theory, called second response
theory, so it propagates quantum systems that are initially described by a general linear combination of
different states, which can include the ground state, and show how with a special set of
time-dependent cluster operators such propagations are performed. Our theory shows strong
consistency with numerically exact results for the determination of quantum mechanical observables,
probabilities, and coherences. We discuss unperturbed non-stationary states within second response
theory and their ability to predict matrix elements that agree with those found in linear and
quadratic response theories. This work also discusses an approximate regularized methodology to
treat systems with potential instabilities in their ground-state cluster amplitudes, and compares
such approximations with respect to reference results from standard unitary theory.
\end{abstract}

\maketitle
\section{Introduction}
The quantitative prediction of the behavior of quantum systems is a critical task in electronic
structure theory and a main motivator to find improved algorithms and approaches for advanced
computing architectures \cite{marzari2021electronic,morton2011theoretical,xie2019insight}, such as quantum computers
\cite{cao2019quantum,mcardle2020quantum,bauer2020quantum,kokcu2022fixed,motta2022emerging}. The field of electronic structure has
advanced significantly in recent years where it is now possible to perform routinary molecular drug
design simulations \cite{raha2007role,cavalli2006target}, materials high-throughput screening
\cite{saal2013materials,greeley2006computational}, simulations of biochemical systems
\cite{van2013combined,jones2020embedding}, among
others. In the field of materials research, for instance, localized states in semiconductors that
arise from impurities can behave as pseudo-atomic states \cite{baskes1992modified} that are excited
by thermal fluctuations or related phenomena. Because of systems like these, a
constant target of theory development focuses on computation of energies and forces. However, other
properties are also crucial for the engineering of quantum systems \cite{lachance2019hybrid}. This
includes variables such as electrical and magnetic multipolar moments, and their related
derivatives. A more advanced type of computation is the direct propagation of a quantum system that
involves excited states in its initial state. This is an important goal since non-linear propagations reveal
behaviors that are inaccessible by the standard analysis of energetic properties. Real time
propagations may also be more computationally efficient than techniques based on
eigenvalues/eigenvectors only
\cite{provorse2016electron,wang2022accelerating}, and offer the advantage of being able to capture
the response of the system to strong and ultra-strong driving fields, as well as their non-adiabatic
couplings to nuclear motion \cite{palacios2020quantum}.

There is a wide set of electronic structure theories available for different types of applications
\cite{samanta2014excited,dral2021molecular,park2020multireference,austin2012quantum,vitillo2022multireference,von2020exploring}.
Among these, coupled cluster (CC) theory is a very reliable formalism to study quantum systems
\cite{sneskov2012excited,zhang2019coupled,liu2021relativistic}. CC methods enjoy the ability
to predict accurate quantities that feature size-extensivity
\cite{bartlett1978many,Paldus1994,bartlett2012coupled,koch1990coupled};
meaning they, in principle, scale correctly to the macroscopic scale. They are systematically
improvable \cite{bartlett2007coupled}, so they are guaranteed to reach exact numerical values if
computational resources allow it. These points have been strong motivators for the scientific
community to explore the potential application of CC methods via quantum computers. Unitary CC theory
\cite{taube2009rethinking,shen2017quantum,harsha2018difference,lee2018generalized,evangelista2019exact},
for example, has gained significant attention for predicting ground-state energies through quantum
algorithms, with promising results \cite{romero2018strategies,xia2020qubit,ryabinkin2018qubit,anand2022quantum}.  
Therefore, it can be expected that, as quantum and standard architectures continue to advance, the
applicability of CC-based theory and algorithms will expand as well.

Ground-state methods in CC theory have reached a highly productive stage for the design and
understanding of molecular systems. Algorithms to study periodic systems have also been deployed recently
\cite{neufeld2022ground,wang2021absorption}. Excited-state methods, on the other hand, are the subject of
on-going theory development. Furthermore, time-dependent (TD) CC formalisms are being
designed and implemented to understand quantum systems beyond equilibrium
\cite{folkestad2023entanglement,sverdrup2023time,skeidsvoll2022simulating,pathak2022time,pathak2021time,
skeidsvoll2020time,pedersen2020interpretation,pedersen2019symplectic,koulias2019relativistic,
white2019time,white2018time,park2019equation,nascimento2017simulation,nascimento2016linear,
pigg2012time,walz2012application,kats2011second,sonk2011td}.
This field started with early theoretical work that explored the potential use of time-dependent CC operators
for the calculation of observable quantities in electronic structure and nuclear dynamics
\cite{monkhorst1977calculation,schonhammer1978time,hoodbhoy1978time,hoodbhoy1979time,arponen1983variational}, and 
it has provided support to such formalisms that continue making progress \cite{sverdrup2023time}.
These have made significant advances for propagations that start from the ground state, and in combination with
non-Hermitian (standard) CC, variational CC, or unitary CC. There are opportunities, however, to
develop formalisms that predict the evolution of general linear combinations of quantum states
(which may or may not include the ground state of the system). As common goals with standard CC
methods, extended TD techniques must be based on linked diagrammatic expressions and systematic
improvability. Quantum superpositions of ground and excited states are fundamental in the study of
coherent phenomena. These take place in molecular and materials excitonics, quantum computers,
ultracold atoms, quantum spectroscopy, nuclear magnetic and electron paramagnetic resonance
experiments, spin lattices, among other physically relevant cases.  For these reasons, theoretical
techniques capable of predicting time-dependent and frequency-space observable signatures of
coherent phenomena are critical. Although we focus in this work on electronic degrees of freedom,
the techniques discussed here can be extended to the study of coherent quantum vibrations, or vibronic states.

Motivated by the points above, a TD CC approach was proposed in Ref. \cite{mosquera2022excited}.
This formalism, called {\slshape second response} (SR) theory, can be used to predict the
time-dependent behavior of systems that are initially in a linear combination of excited states,
while maintaining the desired feature of diagrammatic connectedness (ground states are included
through an extra set of equations). We showed the formalism to be consistent with linear and
quadratic CC response theories, and validated it successfully with model systems
\cite{mosquera2022excited}. SR theory is an extension of previous work focused on the linear
response of excited states \cite{mosquera2016sequential,mosquera2017exciton,mosquera2021second}.
In this work, we show further developments of SR theory in which, by using three TD excitation
vectors, the expectation value of a quantum observable as a function of time can be predicted, and
this includes probabilities and coherences. Our theory propagates general quantum superpositions
that can now include the ground state of the system. We then show the solution of the SR equations
that give cluster operators that describe unperturbed non-stationary states. Finally, we discuss an
approximated method to treat systems that require regularization to achieve stability. These
developments are then verified through a three-electron/three-level system that can be solved
exactly using standard unitary quantum propagations.

\section{Theory}\label{theory}

\subsection{Motivation}\label{motiv}
This work focuses on non-relativistic Coulombic electronic systems.  The free (or unperturbed)
Hamiltonian, denoted as $\hat{H}_0$, is the sum of the kinetic, repulsion, electron-nuclear
attraction (assuming fixed nuclei) components. We are interested in electronic systems that are
subject to an external perturbation, so the TD Hamiltonian of interest reads
$\hat{H}(t)=\hat{H}_0+\hat{V}(t)$, where $\hat{V}(t)$ is the TD perturbing potential operator.  It
is important to distinguish two viewpoints: First, standard TD quantum mechanics provides
expressions and rules to determine a TD observable of interest. We denote the initial state of the
system in the standard quantum mechanical picture as $\Psi(0)$.  Second, our goal is to {\slshape
develop coupled-cluster equations that propagate the CC analogue of} $\Psi(0)$. So throughout this
work we will establish connections between standard quantum mechanical observables and their CC
equivalents.  For convenience we will often use Heisenberg operators: If $\hat{A}$ denotes the
operator corresponding to some observable $A$ of interest, then
$\hat{A}^{\rm{H}}(t)=\hat{U}^{\dagger}(t)\hat{A}\hat{U}(t)$, where
$\hat{U}(t)=\mc{T}\exp[-\rmi\int_0^t \hat{H}(s)\rmd s]$ (with $\mc{T}$ denoting the time-ordering
superoperator).

We define the {\slshape auxiliary wavefunction}:
\ben
|\Psi_{\rm{R}}(t=0;g_{\rm{R}})\rangle = |\Psi_0\rangle+g_{\rm{R}}|\Psi(0)\rangle~,
\een
where:
\ben
|\Psi(0)\rangle=S|\Psi_0\rangle+\sum_{N\neq 0}C_N|\Psi_N\rangle~,
\een
$\Psi_0$ denotes the ground state of the system, $g_{\rm{R}}$ a coefficient, and $\Psi_N$ an excited
state ($N\neq 0$), the sum ($\sum_N$) runs over excited states. We also define the wavefunction:
$\Psi_{\rm{E}}=\sum_{N\neq 0}C_N\Psi_N$. The wavefunctions $\{\Psi_J\}$ satisfy
$\hat{H}_0|\Psi_J\rangle=E_J|\Psi_J\rangle$. $\Psi(0)$ is the initial state of interest to be
propagated. The exact TD wavefunction of the system reads
$|\Psi(t)\rangle=\hat{U}(t)|\Psi(0)\rangle$. If known, $\Psi(t)$ would determine all the measurable
quantities of the system.  Previous work \cite{mosquera2022excited} on SR CC theory showed how to
propagate general arbitrary states by computing separately the CC representations of $\langle
\Psi_{\rm{E}}|\hat{A}^{\rm{H}}(t)|\Psi_0\rangle$, and $\langle
\Psi_{\rm{E}}|\hat{A}^{\rm{H}}(t)|\Psi_{\rm{E}}\rangle$. In this current work we show how, through
an integrated approach, an arbitrary initial state that includes a contribution from the ground
state is propagated in SR CC theory. Both approaches are equivalent and give the same numerical
results, but the present formulation may offer advantages for potential implementations.

SR theory is motivated by the following property of $\Psi_{\rm{R}}$ in standard TD quantum mechanics:
\ben\label{sr_prop_0}
\lim_{g_{\rm{R}}\rightarrow 0}\frac{\partial}{\partial g_{\rm{R}}}
|\Psi_{\rm{R}}(0;g_{\rm{R}})\rangle = |\Psi(0)\rangle~.
\een
The derivative with respect to the coefficient $g_{\rm{R}}$ is equivalent to the wavefunction
$\Psi(0)$, which is a linear combination of quantum states. This combination includes, in principle,
any desired number of eigen-states (with $\{C_N\}$ coefficients), which may include the ground-state wavefunction of the system 
(with $S$ coefficient).

Similarly, we define the conjugate auxiliary wavefunction that depends on a second parameter,
$g_{\rm{L}}$:
\ben\label{sr_prop}
\langle\Psi_{\rm{L}}(0;g_{\rm{L}})| = \langle\Psi_0|+g_{\rm{L}}\langle\Psi(0)|~,
\een
which follows $\lim_{g_{\rm{L}}\rightarrow 0}\partial/\partial g_{\rm{L}}
\langle \Psi_{\rm{L}}(0;g_{\rm{L}})|=\langle \Psi(0)|$.
For CC developments, we denote the operators $\hat{X}^J$ and $\hat{T}$ in
terms of the particle-hole excitation operators $\{\hat{\tau}_{\mu}\}$ (for an excitation labeled
$\mu$, when applied to $|0\rangle$, $\hat{\tau}_{\mu}$ promotes electrons from occupied levels into
virtual ones and returns the corresponding wavefunction determinant, \cite{helgaker2014molecular}), so
$\hat{T}=\sum_{\mu}t_{\mu}\hat{\tau}_{\mu}$ and $\hat{X}^J=\sum_{\mu}X_{\mu}^J\hat{\tau}_{\mu}$.
$\hat{T}$ is the standard ground-state cluster operator; $X^J_{\mu}$ is defined later below.

Even though the limit procedure is redundant for standard linear combinations, for a CC wavefunction
of the form:
\ben
|\Phi(t=0;g_{\rm{R}})\rangle=\exp\Big[\hat{T}+g_{\rm{R}}\Big(S\hat{X}^0+\sum_N
C_N\hat{X}^N\Big)\Big]|0\rangle~,
\een
where $|0\rangle$ is the reference state, it leads to the CC analogue of equation
(\ref{sr_prop_0}), which resembles $\Psi(0)$:
\ben
\lim_{g_{\rm{R}}\rightarrow 0}\frac{\partial}{\partial
g_{\rm{R}}}|\Phi(t=0;g_{\rm{R}})\rangle =
\Big(S\hat{X}^0+\sum_NC_N\hat{X}^N\Big)e^{\hat{T}}|0\rangle~,
\een
where $\hat{X}^N$ is the CC excitation operator from equation-of-motion CC (EOM-CC); $\hat{X}^0$ is
simply the unity in this case \cite{stanton1993equation}.  The derivative and limit then yield a linear
combination of CC wavefunctions, in analogy with the standard quantum mechanical approach.  For
instance, the analogue of the eigen-function $|\Psi_N\rangle$ is the CC wavefunction
$\hat{X}^Ne^{\hat{T}}|0\rangle$. Although these two wavefunctions can in principle lead to the same
excited state energies, they are not exactly the same objects, due to the asymmetric (non-Hermitian)
nature of standard CC theory \cite{koch1990coupled,pedersen1997coupled}. 
For example, the CC analogue of $\langle\Psi_N|$ is $\langle
0|\hat{\Lambda}^N\exp(-\hat{T})$, where $\hat{\Lambda}^N$ is the conjugate excitation operator from
EOM-CC, and it is expanded as
$\hat{\Lambda}^N=\sum_{\mu}\hat{\tau}^{\dagger}_{\mu}\Lambda^N_{\mu}$.  
Therefore, the conjugate version is different (not simply a
Hermitian conjugate) because the ``lambda'' vectors must {\slshape compensate} for the term
$\exp[-\hat{T}-g_{\rm{R}}(S\hat{X}^0+\sum_NC_N\hat{X}^N)]$. Hence we introduce 
the wavefunction (the CC analogue of $\langle \Psi_{\rm{L}}|$):
\ben
\langle \Upsilon(t=0;g_{\rm{L}},g_{\rm{R}})|=\langle
0|\hat{\lambda}_{\rm{m}}(0;g_{\rm{L}},g_{\rm{R}})\exp[-\hat{x}_{\rm{m}}(0;g_{\rm{R}})]~,
\een
where $\hat{x}_{\rm{m}}(0)=\hat{T}+g_{\rm{R}}(S\hat{X}^0+\sum_NC_N\hat{X}^N)$, and
$\hat{\lambda}_{\rm{m}}=\sum_{\mu}\hat{\tau}^{\dagger}_{\mu}\lambda_{\rm{m},\mu}$. 
Our cluster operator expansions here include an identity term, labeled $\mu=0$, where $\hat{\tau}_0=1$.
Due to the non-Hermiticity, the left CC wavefunction has information
about the right-handed problem.  

Returning to the standard quantum mechanical picture, we consider the expectation value of the
auxiliary TD observable $\tilde{A}(t)$:
\ben\label{LRObs}
\tilde{A}(t)=\frac{\langle \Psi_{\rm{L}}(0)|\hat{A}^{\rm{H}}(t)|\Psi_{\rm{R}}(0)\rangle}{\mc{N}^2}~,
\een
where $\mc{N}^2=\langle\Psi_{\rm{L}}(0)|\Psi_{\rm{R}}(0)\rangle=1+g_LS^*+g_RS+g_Lg_R$. This function
satisfies:
\begin{eqnarray}
\partial_{\rm{LR}}^2\tilde{A}(t)&=\lim_{g_{\rm{L}},g_{\rm{R}}\rightarrow 0}\frac{\partial^2}{\partial
g_{\rm{L}}\partial g_{\rm{R}}}\tilde{A}(t)\\
&=\langle \Psi_{\rm{E}}|\hat{A}^{\rm{H}}(t)|\Psi_{\rm{E}}\rangle-(1-|S|^2)\langle
\Psi_0|\hat{A}^{\rm{H}}(t)|\Psi_0\rangle~,\nonumber
\end{eqnarray}
and
\begin{eqnarray}
\partial_{\rm{L}}\tilde{A}(t)&=\lim_{g_{\rm{L}},g_{\rm{R}}\rightarrow 0}\frac{\partial}{\partial
g_{\rm{L}}}\tilde{A}(t)=\langle \Psi_0|\hat{A}^{\rm{H}}(t)|\Psi_{\rm{E}}\rangle~,\\
\partial_{\rm{R}}\tilde{A}(t)&=\lim_{g_{\rm{L}},g_{\rm{R}}\rightarrow 0}\frac{\partial}{\partial
g_{\rm{R}}}\tilde{A}(t)=\langle \Psi_{\rm{E}}|\hat{A}^{\rm{H}}(t)|\Psi_0\rangle~.\nonumber
\end{eqnarray}
The last two equations define the operators
$\partial_{\rm{LR}}^2=\lim_{g_{\rm{L}},g_{\rm{R}}\rightarrow 0} \partial^2/\partial
g_{\rm{L}}\partial g_{\rm{R}}$, $\partial_{\rm{L}}=\lim_{g_{\rm{L}},g_{\rm{R}}\rightarrow 0}\partial/\partial
g_{\rm{L}}$, and $\partial_{\rm{R}}=\lim_{g_{\rm{L}},g_{\rm{R}}\rightarrow 0}\partial/\partial
g_{\rm{R}}$. Using these definitions we obtain that the left CC wavefunction must obey:
\ben
\langle 0|\Big(S^*\hat{L}^0+\sum_NC_N^* \hat{\Lambda}^N\Big)e^{-\hat{T}}=\langle
0|\partial_{\rm{L}}\hat{\lambda}_{\rm{m}}(0)e^{-\hat{T}}~,
\een
giving a term similar to $\langle \Psi(0)|$. We express $\hat{L}_0=1+\hat{\Lambda}$, where
$\hat{\Lambda}$ is the ``lambda'' operator, which is the ground-state left analogue of $\hat{T}$.

From these two expressions we observe that a TD observable of a system that initiates at a
general quantum state, $\Psi(0)$, is obtained (which is a desired result):
\ben\label{psiApsi}
\langle\Psi(t)|\hat{A}|\Psi(t)\rangle =\partial_{\rm{LR}}^2\tilde{A}+
S\partial_{\rm{L}}\tilde{A}+S^*\partial_{\rm{R}}\tilde{A}+
\langle\Psi_0|\hat{A}^{\rm{H}}(t)|\Psi_0\rangle~.
\een
In this work, motivated by this relation, we analyze the auxiliary observable $\tilde{A}$ using
cluster operators and use the results from such analysis to represent
$\langle\Psi(t)|\hat{A}|\Psi(t)\rangle$. In section \ref{res_dis} we apply such
expression to study a model 3-level/3-electron system.

\subsection{Second response theory}\label{second}
The standard TD equations from CC theory are not limited to propagations from the ground state. For
example, using these standard CC equations, one can imagine the situation where a system is driven
from the ground-state to another different quantum state. If the simulation were to be restarted at
such new quantum state, the system is indeed now being propagated from a state that is not the
ground state.  However, the evolved ``right'' excitation TD vector is not connected in a
straightforward fashion to the EOM-CC eigen-vectors, nor the conjugate TD excitation vector. This
motivates our goal, which is to describe the auxiliary observable shown in equation (\ref{LRObs}) through
CC quantities, then use of the {\slshape differential step followed by the limiting procedure}
presented in section \ref{motiv} to compute $\langle \Psi(t)|\hat{A}|\Psi(t)\rangle$ (similarly as
in Ref. \cite{mosquera2022excited}), without calculating the wavefunction $\Psi(t)$ of course. It
allows for a connection between EOM-CC eigen-vectors and general TD propagations, where one sets the
values of the initial-state coefficients $S$, and $\{C_N\}$. For this reason, we examine differentiation
with respect to the superposition parameters $g_{\rm{L}}$ and $g_{\rm{R}}$ of the standard CC motion
equations:
\ben\label{xm_eq}
\ui\partial_t x_{\rm{m},\mu}(t)=\langle \hat{\tau}^{\dagger}_{\mu}
e^{-\hat{x}_{\rm{m}}(t)}\hat{H}(t)e^{+\hat{x}_{\rm{m}}(t)}\rangle_0~,
\een
and
\ben\label{lm_eq}
-\ui\partial_t\lambda_{\rm{m},\mu}(t)=\langle \hat{\lambda}_{\rm{m}}(t)
e^{-\hat{x}_{\rm{m}}(t)}[\hat{H}(t),\hat{\tau}_{\mu}]e^{+\hat{x}_{\rm{m}}(t)}\rangle_0~,
\een
where the subscript ``$\rm{m}$'' indicates the vectors are modified objects as they account for the
new initial conditions (which depend on $g_{\rm{L}}$ and $g_{\rm{R}}$),
$\hat{x}_{\rm{m}}=\sum_{\mu}x_{\rm{m},\mu}\hat{\tau}_{\mu}$, and $\langle \cdot\rangle_0$
denotes average with respect to the reference ($|0\rangle$): $\langle\cdot\rangle_0=\langle 0|\cdot|0\rangle$.
The terms $x_{\rm{m},0}$ and $\lambda_{\rm{m},0}$ are time-independent (this is a consequence of our
TD variational method, \cite{mosquera2022excited}).
The phase follows the relation: $\partial_t\phi(t)=\langle \hat{\lambda}_{\rm{m}}(t)
e^{-\hat{x}_{\rm{m}}(t)}\hat{H}(t)e^{+\hat{x}_{\rm{m}}}(t)\rangle_0$. This number, $\phi(t)$, does
not influence the calculation of observable expectation values, so it will not be considered in
detail in this work (its properties are presented in Ref. \cite{mosquera2022excited}). 
These modified operators determine the TD
behavior of the following ket and its left conjugate:
\begin{eqnarray}
|\Phi(t;g_{\rm{R}})\rangle=\exp[\hat{x}_{\rm{m}}(t)-\ui\phi(t)]|0\rangle~,\\
\langle \Upsilon(t;g_{\rm{R}},g_{\rm{L}})|=\langle
0|\hat{\lambda}_{\rm{m}}(t)\exp[-\hat{x}_{\rm{m}}(t)+\ui\phi(t)]~.
\end{eqnarray}
The inner product of these two CC wavefunctions is $\langle
\Upsilon(t)|\Phi(t)\rangle=\langle\hat{\lambda}_{\rm{m}}(t)\rangle_0$.

As shown in Ref. \cite{mosquera2022excited}, there are four (reduced to three in this work) TD
excitation vectors that are needed to compute TD observables. These are:
$\hat{x}_{\rm{r}}(t)=\partial_{\rm{R}}\hat{x}_{\rm{m}}(t)$,
$\hat{\lambda}_{\rm{l}}(t)=\partial_{\rm{L}}\hat{\lambda}_{\rm{m}}(t)$,
$\hat{\lambda}_{\rm{r}}(t)=\partial_{\rm{R}}\hat{\lambda}_{\rm{m}}(t)$, and
$\hat{\lambda}_{\rm{l,r}}(t)=\partial_{\rm{LR}}^2\hat{\lambda}_{\rm{m}}(t)$. In addition to such
vectors, the excitation vectors for a propagation that starts exclusively from the ground state are required as
well in our theoretical analysis (for this reason our formalism is called ``second response
theory''). They are denoted as $\hat{x}(t)$, and $\hat{\lambda}(t)$, so $\hat{x}(0)=\hat{T}$ and
$\hat{\lambda}(0)=\hat{L}_0$, and they satisfy equations (\ref{xm_eq}) and (\ref{lm_eq}), respectively.
In the limit $g_{\rm{L}}$ and $g_{\rm{R}}$ tend to zero, $\hat{x}_{\rm{m}}(t)$ and
$\hat{\lambda}_{\rm{m}}(t)$ tend to $\hat{x}(t)$ and $\hat{\lambda}(t)$, respectively.  Given
the importance of $\lim_{g_{\rm{R}}\rightarrow 0}\hat{x}_{\rm{m}}(t)=\hat{x}(t)$, we define the quantity:
\ben
\bar{O}_x(t)=e^{-\hat{x}(t)}\hat{O}(t)e^{+\hat{x}(t)}~,
\een
where $\hat{O}(t)$ is some observable such as $\hat{H}(t)$ or $\hat{A}$. The term
$\bar{O}_{x,\tau,\mu}(t)$ is used to denote:
\ben
\bar{O}_{x,\tau,\mu}(t)=[\bar{O}_x(t),\hat{\tau}_{\mu}]~,
\een
and $\bar{O}_T=e^{-\hat{T}}\hat{O}e^{+\hat{T}}$. The $T$-transformed free Hamiltonian is denoted as
$\bar{H}^0_T=e^{-\hat{T}}\hat{H}_0e^{+\hat{T}}$. The Jacobian matrix of EOM-CC is defined as
$\mc{A}_{\mu\nu}=\langle \hat{\tau}^{\dagger}_{\mu}[\bar{H}_T^0,\hat{\tau}_{\nu}]\rangle_0$. In this
way, the vectors $\mb{X}^J$ and $\bm{\Lambda}^J$ satisfy the equation
$\mc{A}\mb{X}^J=\Omega_J\mb{X}^J$ and
$(\bm{\Lambda}^J)^{\rm{T}}\mc{A}=(\bm{\Lambda}^J)^{\rm{T}}\Omega_J$, where
$(\mb{X}^J)_{\mu}=X^J_{\mu}$, $(\bm{\Lambda}^J)_{\mu}=\Lambda^J_{\mu}$, and $\Omega_J$ is the
excitation energy (from ground to excited state, $\Omega_J=E_J-E_0$).

Using the operators defined above and based on section \ref{motiv}, 
the initial conditions for the modified vectors are:
\ben
\hat{x}_{\rm{m}}(0;g_{\rm{R}})=\hat{x}(0)+g_{\rm{R}}\hat{x}_{\rm{r}}(0)~,
\een
and 
\ben\label{llr}
\hat{\lambda}_{\rm{m}}(0;g_{\rm{L}},g_{\rm{R}})=\hat{\lambda}(0)+g_{\rm{L}}\hat{\lambda}_{\rm{l}}(0)+g_{\rm{R}}\hat{\lambda}_{\rm{r}}(0)+g_{\rm{L}}g_{\rm{R}}\hat{\lambda}_{\rm{l,r}}(0)~,
\een
where we will be interested in the limit where $g_{\rm{R}}$ and $g_{\rm{L}}$ tend to zero, as
discussed later on. These limits are necessary to eliminate higher-order terms in powers of
$g_{\rm{R}}$ and $g_{\rm{L}}$ that the modified vectors develop at $t>0$. As mentioned before, the
asymmetric dependencies on $g_{\rm{L}}$ and $g_{\rm{R}}$ by the excitation operators
$\hat{x}_{\rm{m}}(t)$ and $\hat{\lambda}_{\rm{m}}(t)$ are due to the non-Hermitian attributes of the
formalism (these would not take place in unitary CC theory). To elaborate further on this, the
vector $\hat{\lambda}_{\rm{m}}$ has different features than $\hat{x}_{\rm{m}}$ for two reasons:
i), the left expression has information about the right-handed problem; hence there must be a vector
$\hat{\lambda}_{\rm{r}}$. ii), The auxiliary quantum mechanical observable $\tilde{A}(t)$ also
depends on the product
$g_{\rm{L}}g_{\rm{R}}$, but the right CC wavefunction is independent of $g_{\rm{L}}$ and
could not introduce such dependency if $\hat{A}$ were the identity operator. Therefore this
dependency on $g_{\rm{L}}g_{\rm{R}}$ by the CC expression of $\tilde{A}(t)$ requires the term
$\hat{\lambda}_{\rm{l,r}}$, equation (\ref{llr}).

The operators $\hat{x}_{\rm{r}}$, $\hat{\lambda}_{\rm{l}}$, $\hat{\lambda}_{\rm{r}}$, and
$\hat{\lambda}_{\rm{l,r}}$ follow the relations below [obtained from differentiating the motion equations
of $\hat{x}_{\rm{m}}$ and $\hat{\lambda}_{\rm{m}}$, equations (\ref{xm_eq}) and (\ref{lm_eq})] \cite{mosquera2022excited}:
\begin{eqnarray}
&\ui\partial_t x_{\rm{r},\mu}(t)= \langle \hat{\tau}^{\dagger}_{\mu}
[\bar{H}_x(t),\hat{x}_{\rm{r}}(t)]\rangle_0~,\label{all_eqs_a}\\
-&\ui\partial_t \lambda_{\rm{l},\mu}(t)=
\langle \hat{\lambda}_{\rm{l}}(t)
[\bar{H}_x(t),\hat{\tau}_{\mu}]\rangle_0~,\label{all_eqs_b}\\
-&\ui\partial_t \lambda_{\rm{r},\mu}(t)=
\langle \hat{\lambda}(t)[\bar{H}_{x,\tau,\mu}(t),
\hat{x}_{\rm{r}}(t)]
+\hat{\lambda}_{\rm{r}}(t)\bar{H}_{x,\tau,\mu}(t)\rangle_0~,\label{all_eqs_c}\\
-&\ui\partial_t \lambda_{\rm{l,r},\mu}(t)=\langle \hat{\lambda}_{\rm{l,r}}(t)
\bar{H}_{x,\tau,\mu}(t)
+\hat{\lambda}_{\rm{l}}(t)
[\bar{H}_{x,\tau,\mu}(t),\hat{x}_{\rm{r}}(t)]
\rangle_0~\label{all_eqs_d}.
\end{eqnarray}
With the equations developed above one can propagate
the CC analogue of the state $\Psi(0)=S\Psi_0+\sum_NC_N\Psi_N$ (or $\Psi(0)=S\Psi_0+\Psi_{\rm{E}}$). 
This requires that the derivative vectors be expressed in terms of other vectors that ensure
solution to equations (\ref{all_eqs_a}-\ref{all_eqs_d}) and that initially describe a term like
$\langle\Psi(0)|\hat{A}|\Psi(0)\rangle$. First, $\hat{x}_{\rm{r}}(t)$ follows the initial condition:
\ben
\hat{x}_{\rm{r}}(0)=S\hat{X}^0+\sum_NC_N\hat{X}^N~,
\een
and is propagated as expressed in equation (\ref{all_eqs_a}).
Now, to further simplify the solution of these equations, and generalize Ref. \cite{mosquera2022excited} to include a
contribution from the ground state, we write 
\ben\label{leq}
\hat{\lambda}_{\rm{l}}(t)=\hat{\lambda}^{\rm{E}}_{\rm{l}}(t)+S^*\hat{\lambda}(t)~,
\een 
where
$
\hat{\lambda}_{\rm{l}}^{\rm{E}}(0)=\sum_NC_N^*\hat{\Lambda}^N
$, $\hat{\lambda}(0)=\hat{L}_0$, and $t\ge 0$. For the right derivative,
\ben\label{req}
\hat{\lambda}_{\rm{r}}(t)=\hat{\lambda}_{\rm{r}}^{\rm{E}}(t)+S\hat{\lambda}(t)~,
\een 
where
$\hat{\lambda}_{\rm{r}}^{\rm{E}}(0) = -\sum_{N,I}C_NF^{NI}\hat{\Lambda}^I/(\Omega_N+\Omega_I)$,
$t\ge 0$, and
\ben
F^{NI}=\sum_{\mu\nu}\langle\hat{L}_0\big[[\bar{H}_{T}^0,\hat{\tau}_{\mu}],\hat{\tau}_{\nu}\big]\rangle_0
X^N_{\mu}X^I_{\nu}~.
\een
For the term $\hat{\lambda}_{\rm{l,r}}$ we have:
\ben\label{lreq}
\hat{\lambda}_{\rm{l,r}}(t)=\hat{\lambda}_{\rm{l,r}}^{\rm{E}}(t)+S\hat{\lambda}_{\rm{l}}^{\rm{E}}(t)+\hat{\lambda}(t)+S^*\hat{\lambda}_{\rm{r}}^{\rm{E}}(t)~,
\een
where $\hat{\lambda}_{\rm{l,r}}^{\rm{E}}(t=0)=\sum_JY^J\hat{\Lambda}^J$, $t\ge 0$, and
\ben
Y_J=\sum_{N,I}C_N^*C_I\frac{\langle
\hat{\Lambda}^N\big[[\bar{H}^0_T,\hat{X}^J],\hat{X}^I\rangle_0}{\Omega_N-\Omega_J-\Omega_I}~.
\een
This expression and that for $\hat{\lambda}_{\rm{l}}^{\rm{E}}(0)$ are needed 
for the estimation of the excited-state component of the observable
$\tilde{A}(t)$ (as shown in section \ref{unpert}). By substituting
equations (\ref{leq}-\ref{lreq}) into equations (\ref{all_eqs_b}-\ref{all_eqs_d}), we observe that
$\hat{\lambda}_{\rm{l}}^{\rm{E}},~\hat{\lambda}_{\rm{r}}^{\rm{E}}$ and
$\hat{\lambda}_{\rm{l,r}}^{\rm{E}}$ follow the same equations as
$\hat{\lambda}_{\rm{l}},~\hat{\lambda}_{\rm{r}}$ and $\hat{\lambda}_{\rm{l,r}}$ (supplemental material). 
Furthermore, these operators simplify the final expression for a TD observable. 

We now show that the normalized auxiliary observable below provides excited-state information:
\ben
\tilde{A}_{\rm{N}}(t)=\frac{\langle \Upsilon(t)|\hat{A}|\Phi(t)\rangle}{\mc{N}_2^2}~,
\een
where $\mc{N}^2_2=\langle \Upsilon(t)|\Phi(t)\rangle=1+g_{\rm{L}}S^*+g_{\rm{R}}S+g_Lg_R$. The term
$\langle \Upsilon(t)|\hat{A}|\Phi(t)\rangle$ is equivalent to
$\langle\hat{\lambda}_{\rm{m}}(t)\exp[-\hat{x}_{\rm{m}}(t)]\hat{A}\exp[+\hat{x}_{\rm{m}}(t)]\rangle_0$.
This norm is preserved because the relevant scalar amplitudes added to
our CC operators do not evolve as a function of time.  This auxiliary quantity, $\tilde{A}_{\rm{N}}$, is the
CC analogue of the previously defined $\tilde{A}(t)$ TD average.

We observe the following:
\begin{eqnarray}
\partial_{\rm{LR}}^2\tilde{A}_{\rm{N}}(t)&=\langle
\hat{\lambda}_{\rm{l}}(t)[\bar{A}_x(t),\hat{x}_{\rm{r}}(t)]\rangle_0\\
&~~~~-S^*\langle\{\hat{\lambda}(t)
[\bar{A}_x(t),\hat{x}_{\rm{r}}(t)]+\hat{\lambda}_{\rm{r}}(t)\bar{A}_x(t)\}\rangle_0 \nonumber\\
&~~~~+\langle \{\hat{\lambda}_{\rm{l,r}}(t)-S\hat{\lambda}_{\rm{l}}(t)\}\bar{A}_{x}(t)\rangle_0+
(2|S|^2-1)\langle \hat{\lambda}(t) \bar{A}_x(t)\rangle_0~,\nonumber\\
&=\langle
\hat{\lambda}_{\rm{l}}^{\rm{E}}(t)[\bar{A}_x(t),\hat{x}_{\rm{r}}(t)]+
\hat{\lambda}^{\rm{E}}_{\rm{l,r}}(t)\bar{A}_x(t)\rangle_0~.\nonumber
\end{eqnarray}
We then have that:
\begin{eqnarray}
\langle \Psi_{\rm{E}}|A^{\rm{H}}(t)|\Psi_{\rm{E}}\rangle=\langle
\hat{\lambda}_{\rm{l}}^{\rm{E}}&(t)[\bar{A}_x(t),\hat{x}_{\rm{r}}(t)]
+\hat{\lambda}^{\rm{E}}_{\rm{l,r}}(t)\bar{A}_x(t)\rangle_0+\\
&(1-|S|^2)\langle\hat{\lambda}(t)\bar{A}_x(t)\rangle_0~.\nonumber
\end{eqnarray}
The first-order derivatives follow:
\begin{eqnarray}
\partial_{\rm{L}}\tilde{A}_{\rm{N}}(t)&=\langle\hat{\lambda}_{\rm{l}}^{\rm{E}}(t)\bar{A}_x(t)\rangle_0~,\\
\partial_{\rm{R}}\tilde{A}_{\rm{N}}(t)&=\langle\hat{\lambda}_{\rm{r}}^{\rm{E}}(t)\bar{A}_x(t)+\hat{\lambda}(t)[\bar{A}_x(t),\hat{x}_{\rm{r}}(t)]\rangle_0~,
\end{eqnarray}
so $\langle \Psi_0|\hat{A}^{\rm{H}}(t)|\Psi_{\rm{E}}\rangle=\partial_{\rm{R}}\tilde{A}_{\rm{N}}(t)$ and 
$\langle \Psi_{\rm{E}}|\hat{A}^{\rm{H}}(t)|\Psi_0\rangle=\partial_{\rm{L}}\tilde{A}_{\rm{N}}(t)$.
Combining expressions and using equation (\ref{psiApsi}) we obtain the formula:
\ben\label{psiApsi_ii}
\langle \Psi(t)|\hat{A}|\Psi(t)\rangle
=\langle
\hat{\lambda}_{\rm{l}}(t)[\bar{A}_x(t),\hat{x}_{\rm{r}}(t)]+\hat{\lambda}_{\rm{l,r}}(t)\bar{A}_x(t)\rangle_0~.
\een
We also denote the right hand side of the above equation as $\langle \hat{A}\rangle_{\rm{sr}}(t)$,
with the ``sr'' subscript implying the average is computed using the SR cluster operators.  From
this expression we note that in order to predict the evolution of the observable one needs to solve
for three TD cluster operators. This equation above is also more compact than that used in Ref.
\cite{mosquera2022excited}, which would require computing matrix elements separately in the case
where $S\neq 0$.

The relation shown above, equation (\ref{psiApsi_ii}), is not an exact equality, but a rigorous
assignment due to the non-Hermitian nature of the present CC formalism. This aspect is also present
in linear and quadratic CC response theories \cite{pedersen1997coupled}. So a deviation between
standard quantum mechanics and the CC formalism is expected to take place, but in our simulations
this deviation is small, as discussed later. This is an issue that can be solved by restoring
{\slshape Hermiticity and unitarity}, where our theoretical foundation remains applicable.

\subsection{Determining wavefunction probabilities and coherences}

As discussed previously in Ref. \cite{mosquera2022excited}, the CC wavefunctions do not provide
straightforwardly the probability of observing a given eigenstate at a particular point time. In
standard wavefunction theory this is accomplished by projecting the eigenstate on the total
wavefunction and taking square modulus of the result, {\slshape i.e.},
$p_I(t)=|\langle\Psi_I|\Psi(t)\rangle|^2$.  As we show in section \ref{res_dis}, an alternative to
this is to compute such amplitude as a quantum mechanical observable, for example, by defining:
\ben
\hat{\mc{P}}_{IJ} = e^{+\hat{T}}\hat{X}^I|0\rangle\langle 0|\hat{\Lambda}^J e^{-\hat{T}}~,
\een
which is the CC analogue of $|\Psi_I\rangle\langle \Psi_J|$, setting $\hat{A}=\hat{\mc{P}}_{II}$,
and using equation (\ref{psiApsi_ii}) to compute $p_I(t)$. More generally, we define
$\tilde{p}_{IJ}(t)=\langle \hat{\mc{P}}_{IJ}\rangle_{\rm{sr}}(t)$ [that is, taking $\hat{A}$ as
$\hat{\mc{P}}_{IJ}$ in equation (\ref{psiApsi_ii}) and computing accordingly]. The wavefunction-based
coherence matrix element $p_{\rm{wf},IJ}(t)=C_I^*(t)C_J(t)$ has the evident property:
$p_{\rm{wf},IJ}^*(t)=p_{\rm{wf},JI}(t)$.  To satisfy this we compute the coherence as follows:
\begin{eqnarray}\label{pij}
{\rm Re}~p_{IJ}(t) &= \frac{1}{2}\Big[\tilde{p}_{IJ}(t)+\tilde{p}_{JI}(t)\Big]~,\\
{\rm Im}~p_{IJ}(t) &= \frac{1}{2}\Big[\tilde{p}_{IJ}(t)-\tilde{p}_{JI}(t)\Big]~.\nonumber
\end{eqnarray}
In section \ref{res_dis} we compare the complex number $p_{IJ}(t)$ to the
object $p_{\rm{wf},IJ}(t)$ defined above.

\section{Unperturbed non-stationary states}\label{unpert}
In this section we show that well-known matrix elements from linear and quadratic response CC theory can be
obtained on the basis of unperturbed SR non-stationary states. We present this development as an
example of the consistency of SR theory with respect to LR and QR theory results.

The initial conditions of the excitation vectors can be derived under two considerations: i), The
SR theory averages must predict very accurately the exact unitary theory results, and, ii), if the system
was originally described by an excited state wavefunction, it should remain in that state when no
external perturbation is applied. This is known as a stationary state. Therefore, we need to examine
solutions to the set shown in equations (\ref{all_eqs_a}-\ref{all_eqs_d}) is absence of
perturbations, so $\bar{H}_x(t)=\bar{H}^0_T$. For
$\hat{x}_{\rm{r}}$ and $\hat{\lambda}_{\rm{l}}$, we have that:
$\hat{x}_{\rm{r}}(t)=\hat{X}^N\exp(-\ui\Omega_N t)$ and
$\hat{\lambda}_{\rm{l}}(t)=\hat{\Lambda}^N\exp(+\ui\Omega_N t)$.
These ensure solution to the Jacobian eigenvalue problem: $\ui\partial_t
\mb{x}_{\rm{r}}=\Omega_N\mb{x}_{\rm{r}}=\mc{A}\mb{x}_{\rm{r}}$, and
$-\ui\partial_t\bm{\lambda}_{\rm{l}}^{\rm{T}}=\Omega_N\bm{\lambda}_{\rm{l}}^{\rm{T}}=\bm{\lambda}^{\rm{T}}_{\rm{l}}\mc{A}$
(where ``$\rm{T}$'' denotes transpose operation).

Now suppose the system is described by a linear superposition of states, leading to an unperturbed
non-stationary state, so
\begin{eqnarray}
\hat{x}_{\rm{r}}(t)&=S\hat{X}^0+\sum_N C_N\hat{X}^N\exp(-\ui\Omega_N t)~,\\
\hat{\lambda}_{\rm{l}}(t)&=S^*\hat{L}_0+\sum_N C_N^*\hat{\Lambda}^N\exp(+\ui\Omega_N t)~.
\end{eqnarray}
With the two solutions above, we can find $\hat{\lambda}_{\rm{r}}$ and $\hat{\lambda}_{\rm{l,r}}$.
First, we express $\hat{\lambda}_{\rm{r}}(t)=\sum_J d_J(t)\hat{\Lambda}^J+S\hat{L}_0$, where $\{d_J\}$ are TD
coefficients (from section \ref{second} we note that $\hat{\lambda}(t)=\hat{L}_0$ and
$\hat{x}(t)=\hat{T}$ in absence of
external perturbations). By using the bi-orthogonality property of the excitation vectors
$\{\hat{X}^J,\hat{\Lambda}^J\}$ we observe that 
\ben
-\ui\partial_t d_J=\sum_N C_NF^{JN}e^{-\ui\Omega_Nt}+\Omega_J d_J(t)~,
\een
so $d_J(t)=-\sum_N C_NF^{NJ}\exp(-\ui\Omega_N t)/(\Omega_N+\Omega_J)$. This gives:
\ben
\hat{\lambda}_{\rm{r}}^{\rm{E}}(t)=-\sum_{N,J}\frac{F^{NJ}}{\Omega_N+\Omega_J}\hat{\Lambda}^Je^{-\ui\Omega_N
t}~,
\een
which in turn leads to the TD dependency of $\hat{\lambda}_{\rm{r}}$, that is,
$\hat{\lambda}_{\rm{r}}(t)=\hat{\lambda}^{\rm{E}}_{\rm{r}}(t)+S\hat{L}_0$.
In a similar fashion, if we write
$\hat{\lambda}_{\rm{l,r}}(t)=\sum_Jy_J(t)\hat{\Lambda}^J+S\hat{\lambda}_{\rm{l}}(t)+
\hat{\lambda}(t)+S^*\hat{\lambda}^{\rm{E}}_{\rm{r}}(t)$, we have
that 
\begin{eqnarray}
-\ui\partial_t &y_J=\\\nonumber
&\Omega_Jy_J(t)+\sum_{M,N}C^*_MC_N\exp[-\ui(\Omega_N-\Omega_M)t]\langle
\hat{\Lambda}^M\big[[\bar{H}^0_T,\hat{X}^J],\hat{X}^N\big]\rangle_0~,
\end{eqnarray}
where the sums run over excited states. The solution to the above equation is:
\ben
y_J(t)=\sum_{M,N}C^*_MC_N\exp[-\ui(\Omega_N-\Omega_M)t]\frac{\langle
\hat{\Lambda}^M\big[[\bar{H}^0_T,\hat{X}^J],\hat{X}^N\big]\rangle_0}{\Omega_M-\Omega_J-\Omega_N}~.
\een
Using the last relation we have that $Y_J=y_J(t=0)$. 

The expressions above for perturbation-free evolutions lead to results consistent with linear and
quadratic CC response theories. Now, suppose the TD
wavefunction is given by $\Psi(t)=\exp(-\ui E_0 t)[S\Psi_0+\sum_NC_N\Psi_N\exp(-\ui\Omega_Nt)]$. We
note that:
\begin{eqnarray}
\langle \Psi(t)|\hat{A}|\Psi(t)\rangle=|S|^2&\langle\Psi_0|\hat{A}|\Psi_0\rangle
+\Big[S^*\sum_N e^{-\ui\Omega_N t}\langle\Psi_0|\hat{A}|\Psi_N\rangle+\rm{c.c.}\Big]\\
&+\sum_{M,N}C_M^*C_Ne^{-\ui(\Omega_N-\Omega_M) t}\langle \Psi_M|\hat{A}|\Psi_N\rangle~.
\end{eqnarray}
Examining the CC analogue of the above equation, based on common phasor terms, we identify the
required matrix elements. These are, the term
\ben\label{te0}
\langle \Psi_N|\hat{A}|\Psi_0\rangle=\langle
\hat{\Lambda}^N\bar{A}_T\rangle_0~,
\een
its conjugate:
\ben\label{te1}
\langle
\Psi_0|\hat{A}|\Psi_N\rangle=\langle\hat{L}_0[\bar{A}_T,\hat{X}^N]\rangle_0-\sum_J\frac{F^{NJ}}{\Omega_J+\Omega_N}\langle
\hat{\Lambda}^J\bar{A}_T\rangle_0~,
\een
and the excited-state/excited-state element:
\begin{eqnarray}
\langle \Psi_M|\hat{A}|\Psi_N\rangle=\delta_{MN}&\langle\hat{L}_0\bar{A}_T\rangle_0+
\langle \hat{\Lambda}^M[\bar{A}_T,\hat{X}^N]\rangle_0\\\label{qr}
&+\sum_J \frac{\langle
\hat{\Lambda}^M\big[[\bar{H}^0_T,\hat{X}^J],\hat{X}^N\big]\rangle_0}{\Omega_M-\Omega_J-\Omega_N}\langle
\hat{\Lambda}^J\bar{A}_T\rangle_0~.
\end{eqnarray}
Equations (\ref{te0}) and (\ref{te1}) are identical to those from CC linear response theory, and
equation (\ref{qr}) was derived originally within the context of CC quadratic response theory \cite{koch1990coupled}. These
equations clearly illustrate the non-Hermiticity of the present formalism, where interchanging
labels ($N\leftrightarrow M$, for example) gives different formulas \cite{pedersen1997coupled}.

\section{Approximated scheme based on regularization}
Even though standard CC theory is designed to address dynamically correlated systems, or more
challenging cases through spin symmetry breaking, there can be scenarios where solving the CC
equations is difficult due to low-energy gaps or related instabilities in the numerical solution
methods used to determine $\{t_{\mu}\}$ (these can occur in perturbation theory applications
\cite{lawler2008penalty,stuck2013regularized}). The CC variational method is not necessarily
restricted in use to non-relativistic Coulombic Hamiltonians, but it can also be applied to
double-hybrid DFT Hamiltonians, and model systems such as those described by Hubbard Hamiltonians
(or similar), where this issue may occur. To circumvent numerical instabilities ({\slshape i.e.},
divergences in $\{t_{\mu}\}$) when they take place
and facilitate the determination of a wavefunction of the form $\exp(\hat{T})|0\rangle$,
regularization can be an appealing approach. It may not provide a completely physically meaningful
wavefunction in some cases, but such wavefunction could be used as a starting point for alternative
formalisms that could need it. In the context of DFT, we showed for the dissociation of fundamental
diatomic molecules that regularization can eliminate instabilities and give results with consistent
physical meaning \cite{jacobson2023cluster}.

As we proposed it previously \cite{mosquera2021density}, one can regularize the equations by adding
an additional term, and find the stationary point of the function (inspired from the standard
variational method \cite{helgaker1988analytical,helgaker1989numerically}):
\ben
\mc{E}(\alpha)=\underset{\bm{\Lambda}, \mb{t}}{\rm{stat.}} \Big[\langle (1+\hat{\Lambda})e^{-\hat{T}}\hat{H}_0
e^{+\hat{T}}\rangle_0+\alpha\bm{\Lambda}\cdot\mb{t}\Big]~,
\een
where $\alpha$ is a regularization number, $\bm{\Lambda}\cdot\mb{t}=\sum_{\mu}\Lambda_{\mu}t_{\mu}$,
and ``$\rm{stat.}$'' stands for stationarization (the minimization principle does not apply to this
case). We denote the solution to the problem above as $\{\Lambda_{\mu}'\}$ and $\{t_{\mu}'\}$ (the
prime symbol indicating these are regularized solutions). We also define
$\hat{L}'=1+\hat{\Lambda}'$. In this section we focus on the eigen-value problem.

We wish to apply the SR method in this case for $t=0$ only. With the solution regularized, the wavefunction
$\Psi_{\rm{R}}$ takes a different form, where the term that is not multiplied by $g_{\rm{R}}$ is a
reference state that is the standard quantum mechanical analogue of $\exp(\hat{T}')|0\rangle$ (which
is not the true ground state CC wavefunction). If we denote this wavefunction as $\Psi_0'$, then
$
|\Psi_R(0,g_R)\rangle=|\Psi_0'\rangle+g_{\rm{R}}|\Psi(0)\rangle,
$
where $\Psi(0)$ is a state we want to represent using a CC object, such as the true ground
state or an excited state of interest. The CC wavefunctions that we
seek here are of the form $\hat{X}^N\exp(\hat{T}')|0\rangle$ and $\langle
0|\exp(-\hat{T}')\hat{\Lambda}^N$. So we take $\hat{x}(0)=\hat{T}'$ and $\lambda(0)=\hat{L}'$.

For $\hat{A}=\hat{H}_0$, we then have, from equation (\ref{psiApsi_ii}), the approximation:
\ben
E = \langle \hat{\Lambda}[\bar{H}^0_{T'},\hat{X}]\rangle_0 + \langle
\hat{L}'\bar{H}^0_{T'}\rangle_0~,
\een
where $E$ represents $\langle \Psi(0)|\hat{H}|\Psi(0)\rangle$, and $\bar{H}^0_{T'}=
e^{-\hat{T}'}\hat{H}_0e^{+\hat{T}'}$. We treat $\hat{X}$ and $\hat{\Lambda}$ as variational
quantities. We also define
$
\tilde{\mc{A}}_{\mu\nu}=\langle \hat{\tau}^{\dagger}_{\mu}[\bar{H}_{T'},\hat{\tau}_{\nu}]\rangle_0,
$
which is now based on the regularized Hamiltonian. Note that the identity operator $\hat{\tau}_0=1$
is not required to expand the excitation operators ($\hat{X}$ and $\hat{\Lambda}$) because it
commutes with $\bar{H}_{T'}$ (but in a different formulation including $\hat{\tau}_0$ may be necessary).

With these definitions we have:
\ben
E=(\bm{\Lambda})^{\rm{T}}\tilde{\mc{A}}\mb{X}+\langle \hat{L}'\bar{H}^0_{T'}\rangle_0~,
\een
and the eigenvalue problem takes the form:
$(\bm{\Lambda}^N)^{\rm{T}}\tilde{\mc{A}}=(\bm{\Lambda}^N)^{\rm{T}}\tilde{\Omega}_N$ and
$\tilde{\mc{A}}\mb{X}^N=\tilde{\Omega}_N\mb{X}^N$, where $\tilde{\Omega}_{I}=E_I-E'_0$ and
$E'_0=\langle \hat{L}'\bar{H}^0_{T'}\rangle_0$. The estimated energies of interest are then
$E_I=\tilde{\Omega}_I+E'_0$. As expected, by solving this problem one also obtains approximated values for
the excited-state energies ($\{E_I\},~I\neq 0$).

Assuming that $\hat{\lambda}(0)=\hat{L}'$ renders the present method as approximated in essence. It
can be shown that in a more rigorous approach $\hat{\lambda}(0)$ involves more terms, but this
requires a careful response theory analysis that includes the fact that $\exp(\hat{T}')|0\rangle$ is a linear
combination of the eigen-states of the Hamiltonian $\hat{H}_0$, which is beyond the current scope.


\section{Computational methodology}\label{comput}

\begin{figure}[htb]
\centering
\includegraphics[scale=0.6]{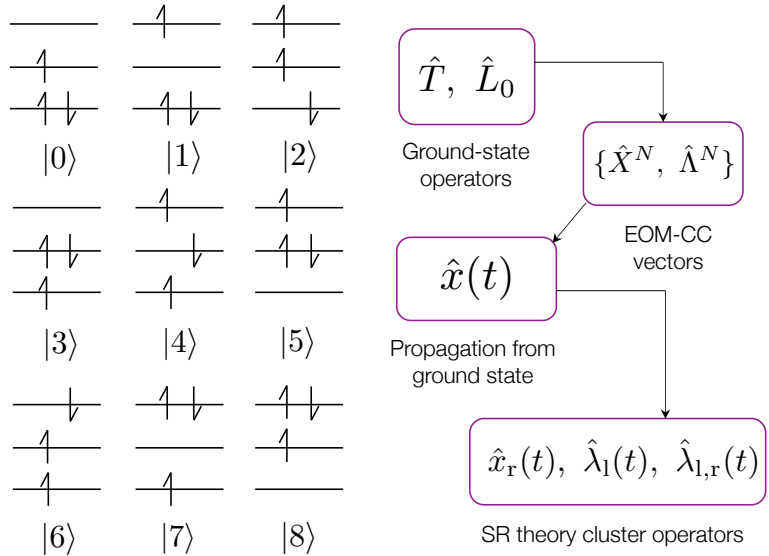}
\caption{Electronic configurations used (left), and operators needed to compute properties
arising from propagations that start from a quantum linear superposition (right). A quantum mechanical
TD average is then computed using the vectors shown in this figure, and equation (\ref{psiApsi_ii}).
The reference is the state $|0\rangle$.}\label{scheme}
\end{figure}

To explore the SR CC theoretical propagation method developed in this work, we examine a
three-level/three-electron system. In the reference configuration $|0\rangle$, the first energy level, labeled
``$j$'', is doubly occupied, the next level, ``$i$'', is singly occupied, and the virtual level,
``$a$'', is unoccupied. We consider a non-trivial Hamiltonian that features two-electron
interactions, and a strong driving perturbation. The unperturbed Hamiltonian has the
form:
\ben
\hat{H}_0=\hat{h}_0+\sum_{q}U_q\hat{n}_{q\uparrow}\hat{n}_{q\downarrow}
+\Big[\sum_{\mu>\nu}W_{\mu\nu}\hat{\tau}_{\mu}\hat{\tau}_{\nu}+\rm{H.c.}\Big]~,
\een
where the single particle Hamiltonian, $\hat{h}_0$, reads:
$\hat{h}_0=\sum_p\epsilon_p\hat{n}_p+\sum_{\mu}b_{\mu}[\hat{\tau}_{\mu}+\hat{\tau}^{\dagger}_{\mu}]$.
The number operator for a spin-level reads
$\hat{n}_{q\sigma}=\hat{q}^{\dagger}_{\sigma}\hat{q}_{\sigma}$, and
$\hat{n}_p=\hat{n}_{p\downarrow}+\hat{n}_{p\uparrow}$ is the number operator of level $p$. We take
$b_{\mu}=b$, $W_{\mu\nu}=w_0$, $U_a=U_i=U_0$, and $U_j=0$.  All the single-particle orbital levels are equally
spaced by a $\Delta \epsilon$ term, and we also set $\epsilon_j=0$ for convenience.

Having defined the free Hamiltonian of the system, the full TD energy operator is defined as:
\ben
\hat{H}(t)=\hat{H}_0-f(t)\hat{D}~.
\een
In this case we take the dipole operator as
$\hat{D}=d_0\sum_{\mu}[\hat{\tau}_{\mu}+\hat{\tau}^{\dagger}_{\mu}]$, and the driving function
corresponds to a rectangular pulse of the form $f(t)=f_0$ if $0<t<t_0$ ($t_0=5~\rm{fs}$), 
$f(t)=0$, otherwise. 

In section \ref{res_dis} we propagate standard and CC-based quantum superpositions and compare the
results. We then show that the regularized scheme leads to consistent results with respect to both the
reference and the SR CC results. For these numerical simulations we assume that $\Delta \epsilon =
1~\rm{eV}$, $\epsilon_j = 0$, $\epsilon_i=\Delta\epsilon$, and $\epsilon_a=2\Delta \epsilon$. We
take $b=0.1~\rm{eV}$, $w_0=U_0=0.2~\rm{eV}$, $d_0=0.5$, and $f_0=0.04~\rm{au}$ (or $2.1\times
10^{10}~{\rm V/m}$); other values
are specified accordingly in section \ref{res_dis}. Figure \ref{scheme} shows a basic schematics of
the three-electron/three-level system of interest in terms of its single-particle basis. There are
nine configurations that lead to nine full-body wavefunctions.  

In physical terms, an $M_{\rm{s}}=+1/2$ system with equally ``spaced'' energy levels, initially in a
linear combination of states, is subject to an external perturbation.  We use our SR CC method to
compute expectation values and compare against numerically exact standard (unitary) quantum
mechanics. In figure \ref{scheme} we show the steps required to perform a propagation based on SR
theory. First, one solves the ground- and excited-state CC problems, which give the operators
$\hat{T}$, $\hat{\Lambda}$, $\{\hat{X}^N\}$, and $\{\hat{\Lambda}^N\}$. Then one determines the
operator $\hat{x}(t)$, through solution of the standard TD CC equations that propagate a system that
initiates at the ground-state. Following this step, one solves the SR TD relations, equations
(\ref{all_eqs_a}, \ref{all_eqs_b}, and \ref{all_eqs_d}), for $\hat{x}_{\rm{r}}$,
$\hat{\lambda}_{\rm{l}}$, and $\hat{\lambda}_{\rm{l,r}}$, and uses these vectors in equation
(\ref{psiApsi_ii}) to compute the TD observable $A$. The initial conditions are applied as well, as
discussed in section \ref{second}, and which are based on the $S$, and $\{C_N\}$ coefficients. To solve the
TD SR equations, we use the midpoint method (or second order Runge-Kutta technique) in the range
$[0-50~\rm{fs}]$; this is based on 50,000 steps. 

\section{Results and discussion}\label{res_dis}

\begin{figure}[htb!]
\centering
\includegraphics[scale=0.5]{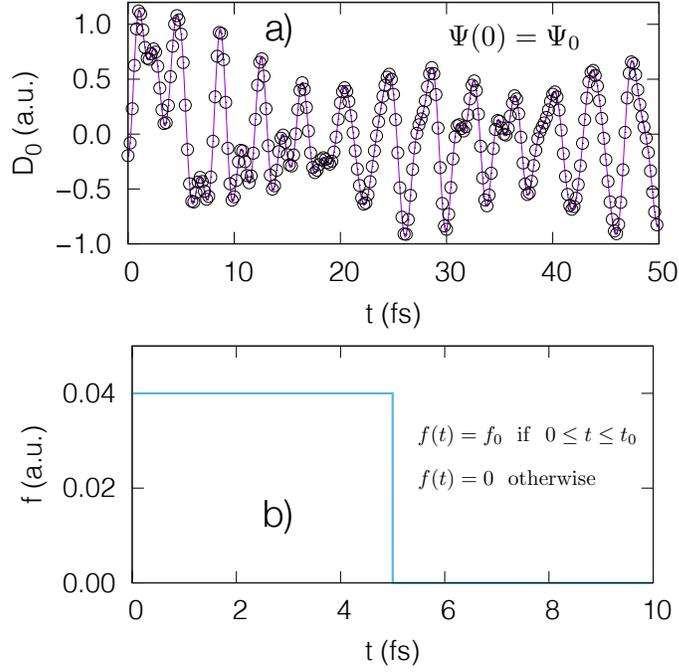}
\caption{Standard propagation and driving pulse. a), Time-dependent dipole,
$D_0(t)=\langle\Psi_0|\hat{D}^{\rm{H}}(t)|\Psi_0\rangle$, for propagation from the ground-state of
the system: open circles denote the standard TD CC propagation, purple line refers to the unitary
reference result (numerically exact). b), Shape of the pulse applied (shown between 0 and 10 fs),
which is turned-off at $t=5~\rm{fs}$.}\label{ah00}
\end{figure}

The standard eigenstates of the system are obtained by diagonalizing the unperturbed Hamiltonian,
such that $\hat{H}_0|\Psi_J\rangle = E_J|\Psi_J\rangle$, $J=0,1,\ldots, 8$ (details about these
states can be found in the supplemental material). In an analogous fashion, we first solve the
ground-state CC problem iteratively to find $\hat{T}$ and $\hat{\Lambda}$, then the left and right
eigenvalue problems (with matrix $\mc{A}$) are solved to determine the set
$\{\mb{\Lambda}^I,~\mb{X}^I\}$ ($I=1,\ldots,8$). Having these eigenvalue problems solved, in order
to generate reference quantum mechanical data, we proceed to propagate the standard wavefunction of
the system using $\ui\partial_t\Psi(t)=\hat{H}(t)\Psi(t)$ [as $\Psi(t+\delta t)=\exp\{-\ui
\delta t\hat{H}(t+\delta t/2)\}\Psi(t)$]. Then, we solve the second response
equations for $\hat{\lambda}_{\rm{l}}$, $\hat{\lambda}_{\rm{l,r}}$, and $\hat{x}_{\rm{r}}$, equations
(\ref{all_eqs_a}, \ref{all_eqs_b}, and \ref{all_eqs_d}), 
as mentioned in section \ref{comput}. The TD observable is computed for both cases
using $\langle\Psi(0)|\hat{A}^{\rm{H}}(t)|\Psi(0)\rangle$ (which serves as reference) and $\langle
\hat{\lambda}_{\rm{l}}(t)[\bar{A}_{x}(t),\hat{x}_{\rm{r}}(t)]+
\hat{\lambda}_{\rm{l,r}}(t)\bar{A}_{x}(t)\rangle_0$. We study the propagation of three different
initial states, denoted ``QS1'', ``QS2'', and ``QS3'', respectively (``QS'' standing for ``quantum
superposition''). These initial states are
defined as follows:
\begin{eqnarray}\label{initial_eqs}
\rm{QS1}&:~~\Psi(0)=\frac{1}{\sqrt{3}}\Psi_0+\frac{1}{\sqrt{3}}\Psi_1+\frac{1}{\sqrt{3}}\Psi_2~,\\
\rm{QS2}&:~~\Psi(0)=\frac{1}{\sqrt{2}}\Psi_7+\frac{\rmi}{\sqrt{2}}\Psi_8~,\\
\rm{QS3}&:~~\Psi(0)=\frac{1}{\sqrt{4}}\Psi_0+\frac{1}{\sqrt{4}}\Psi_3+\frac{\rm{1}}{\sqrt{2}}\Psi_5~.
\end{eqnarray}
Note that QS2 features a complex-valued coefficient ($C_8=i/\sqrt{2}$).

\begin{figure}[htb!]
\centering
\includegraphics[scale=0.5]{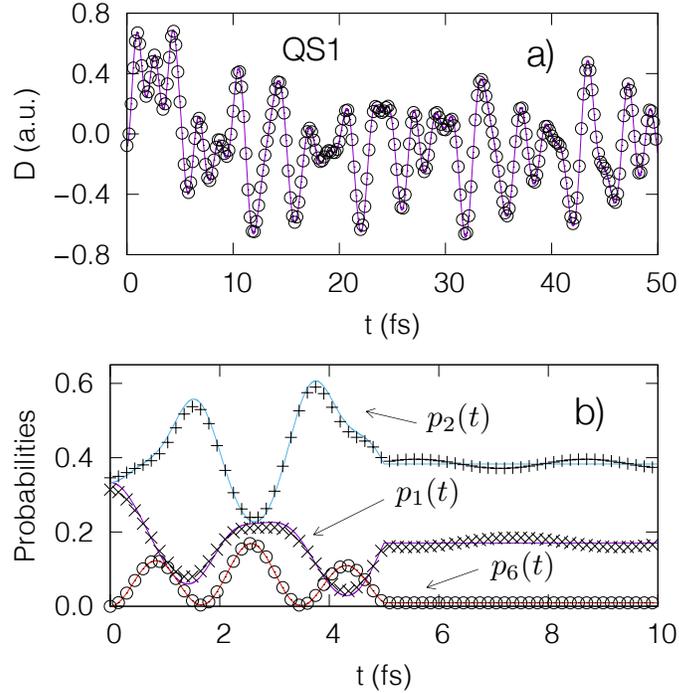}
\caption{Evolution of the system as initiating in the QS1 state in terms of the dipole and
probabilities, $D(t)=\langle{\rm QS}1|\hat{D}^{\rm{H}}(t)|{\rm QS}1\rangle$, where $|\rm{QS1}\rangle$
denotes the first initial quantum superposition shown in equation (\ref{initial_eqs}). a), TD
dipole: purple line corresponds to the numerically exact unitary result, open circles denote the SR
CC theory results. b), Probabilities of the unperturbed Hamiltonian eigenstates $\Psi_1$ (purple
line: exact result, cross symbol: SR theory), $\Psi_2$ (blue line: exact result, ``+'' symbol: SR
theory), and $\Psi_6$ (red line: unitary simulation, open circle: SR theory).}\label{sr_qs1}
\end{figure}

\begin{figure}[htb!]
\centering
\includegraphics[scale=0.5]{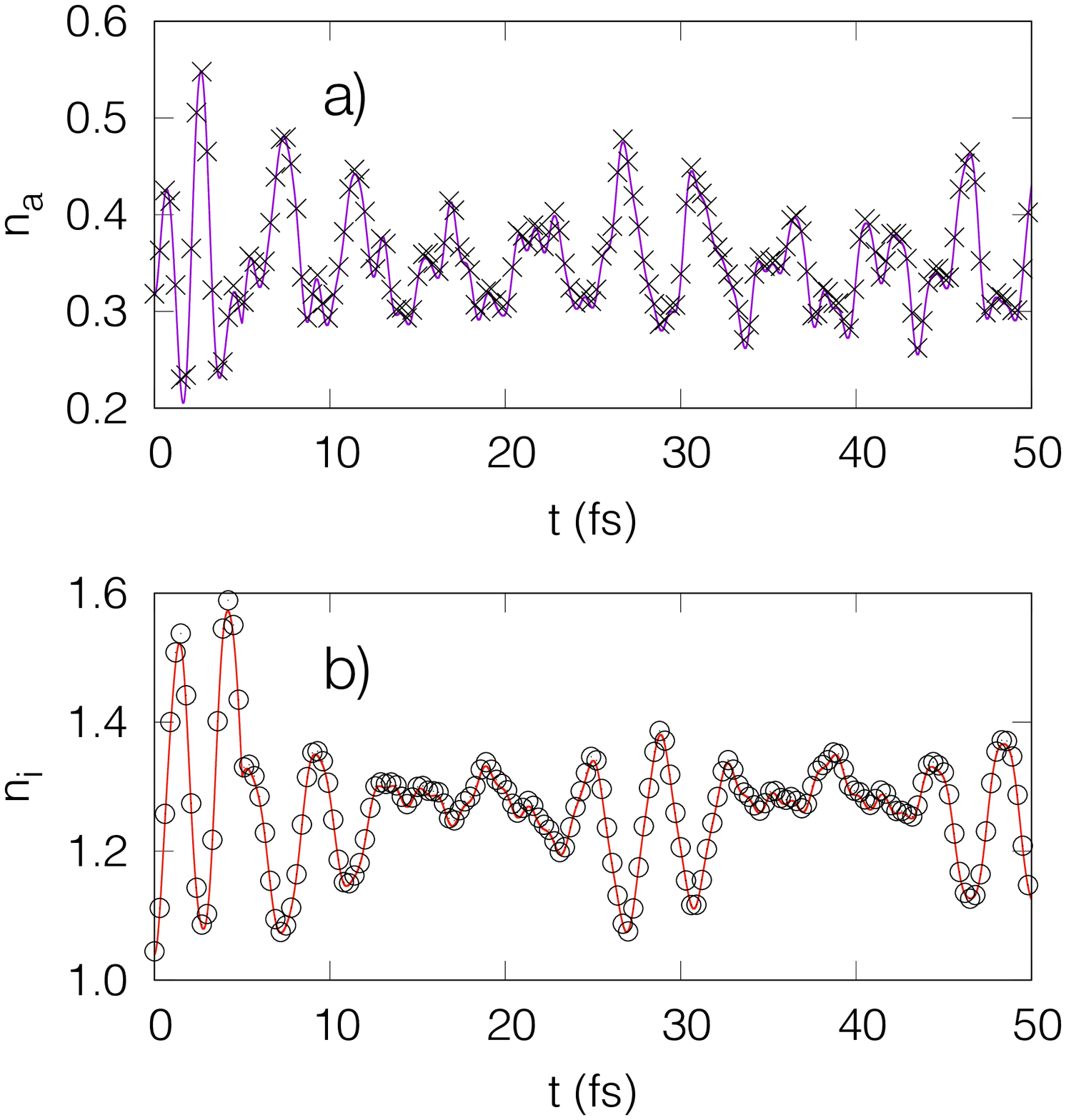}
\caption{a), TD population of level $a$,
purple line: exact result, cross symbols: SR theory. b), TD population of level $i$, red line:
unitary simulation, open circles: SR CC theory.}\label{pops_qs1}
\end{figure}

The Hamiltonian $\hat{H}_0$ yields a ground state that is non-trivial, where the contributions of
the singles and doubles configurations are in the range $[0.04-0.1]$. These are significant
contributions to the reference of the system. Figure \ref{ah00}.a shows the result of propagating
the {\slshape ground-state} using the standard unitary and the CC methods. As expected, the TD
standard CC equations reproduce quite well the unitary results. As mentioned before, from this
propagation one needs the vector $\hat{x}(t)$ to propagate quantum superpositions; the operator
$\hat{\lambda}(t)$ is not needed as $\hat{\lambda}_{\rm{l,r}}(t)$ contains the information about
this cluster operator. To illustrate the applicability of the SR method, we begin considering the
initial state ``QS1'', $\Psi(0)=C_0\Psi_0+C_1\Psi_1+C_2\Psi_2$, where $C_0=C_1=C_2=1/\sqrt{3}$ (also
note that $S=C_0$). The
state $\Psi_0$ is dominated by the reference configuration ($|0\rangle$), $\Psi_1$ by the first
configuration ($|1\rangle$), and $\Psi_2$ by the third configuration ($|3\rangle$); this also
applies to the their corresponding EOM-CC eigen-states. Figure \ref{sr_qs1}.a displays the
comparison between the standard unitary and SR theory propagations for the dipole, and figure
\ref{sr_qs1}.b the eigenstate probabilities.  The agreement between these two propagations is
noticeable. Our propagations are entirely coherent due to the absence of dissipative effects. For
this reason, after the perturbation is turned off, the dipole remains oscillating. The exact
probabilities, as expected, reach a steady value after $t=5~\rm{fs}$. The estimates obtained from
the CC operator $\hat{\mc{P}}_{II}$, except $p_6(t)$, however, show relatively small deviations
after $5~\rm{fs}$. The use of this operator, nonetheless, yields a more desirable consistency than
the other estimators discussed previously in Ref. \cite{mosquera2022excited}. We attribute this
again to the asymmetric characteristics of the formalism and the truncation (up to the fourth order)
of $\exp[-\hat{x}(t)]\hat{\mc{P}}_{IJ}\exp[\hat{x}(t)]$. Figure \ref{pops_qs1} displays the TD
evolution of the populations of the levels $i$ and $a$, where a better overall agreement is observed
in comparison to the eigenstate probability case.

\begin{figure}[htb!]
\centering
\includegraphics[scale=0.5]{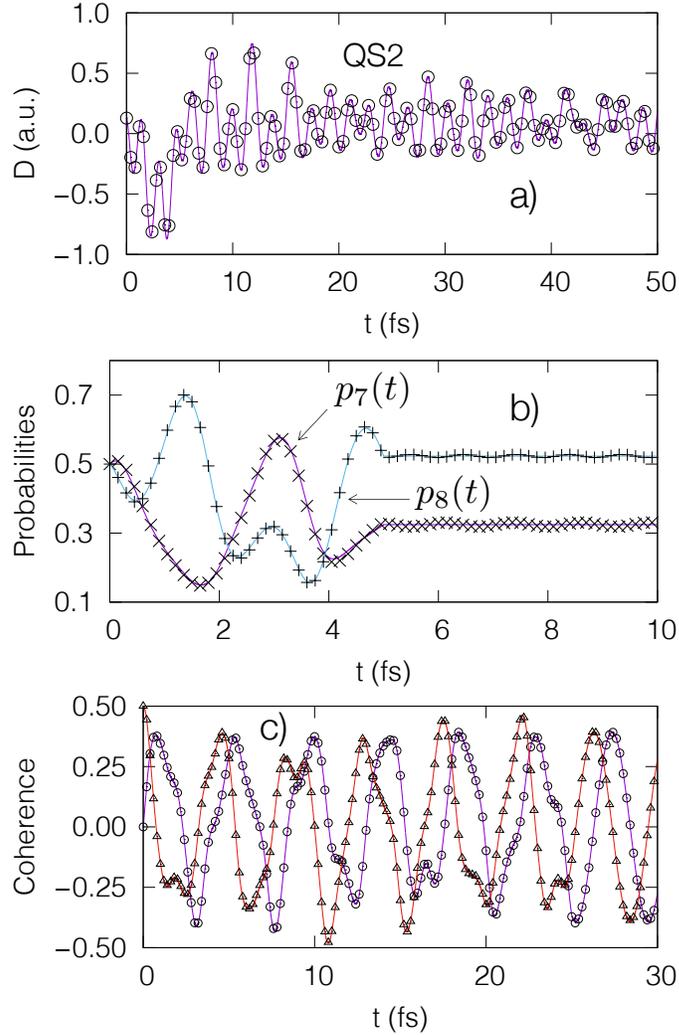}
\caption{Propagation initiating at the QS2 state. a), TD dipole of the system. Purple line
denotes the unitary reference simulation, open circles refer to the SR theory results. b), TD
populations of eigenstates $\Psi_7$ (purple line: exact result, cross symbol: SR simulation) and
$\Psi_8$ (blue line: unitary reference result, ``+'' symbol: SR theory). c), Coherence of states 7
and 8 (or $p_{78}(t)$) as a function of time; red line: exact result for the real part of this
coherence, open triangle: SR theory for the real part; purple line: exact result for imaginary
component, open circle: SR theory for this imaginary part. }\label{qs2}
\end{figure}

A similar consistent behavior between SR CC and unitary results holds
for the other two quantum superpositions, QS2 and QS3. For propagation starting
at QS2, in figure \ref{qs2} we can note the excellent agreement for the
TD dipole and probabilities. QS2 is a 50 \%-50 \% combination of the last two excited states, 
$\Psi_7$ and $\Psi_8$. $\Psi_7$ is dominated by the 5th and 7th configurations, whereas $\Psi_8$ is dominated
by the 8th configuration; these are doubles determinants. In figure \ref{qs2}.a we observe that the oscillating patterns of the
dipole change after the external field is turned off. Even though the field is on for a short period
time, this range of time ($0-5~\rm{fs}$) is sufficient to drive the eigenstate TD probabilities in a non-trivial
fashion, as evidenced in figure \ref{qs2}, where a strong population change occurs. Simulations
where the external field is deactivated at longer times were also explored in our present studies without
noticing differences in deviations with respect to those presented in this section.  On the
other hand, the coherence term $p_{\rm{wf},78}=C_7^*(t)C_8(t)$ is also described with a striking
agreement by the CC operator $\hat{\mc{P}}_{78}$ in conjunction with equation (\ref{pij}), figure
\ref{qs2}.c. These results suggest the SR formalism offers a promising theoretical method to propagate quantum superpositions
while preserving connectedness of quantum mechanical expressions, a condition required to approach
macroscopic/periodic systems; where our developments, besides expanding Ref.
\cite{mosquera2022excited} with the compact inclusion of general quantum superpositions, also provide tools to
explore non-diagonal elements of the quantum mechanical density matrix, which depends on an operator
such as $\hat{\mc{P}}_{IJ}$.

In our simulations, the operator $\hat{\mc{P}}_{IJ}$ was expressed in matrix form by explicitly
computing the exponential matrices involved. The resulting matrix was treated as a matrix associated
to an observable, which was then symmetrized. In more practical settings, however, computationally
efficient ways to use the operator $\hat{\mc{P}}_{IJ}$ would be needed. This can be carried out by
examining the expressions that result from inserting this operator in equation \ref{psiApsi}, where
a considerable aspect to address in future work concerns truncation of the exponential operators.
Furthermore, the symmetrization step is not unique, as expected, as other procedures are possible,
and the trace is not strictly the unity if the exponential operator is computed as a truncated
power series; otherwise the approach is norm-conserving. In a unitary version of SR CC, however,
this issue may not arise.

\begin{figure}[htb!]
\centering
\includegraphics[scale=0.5]{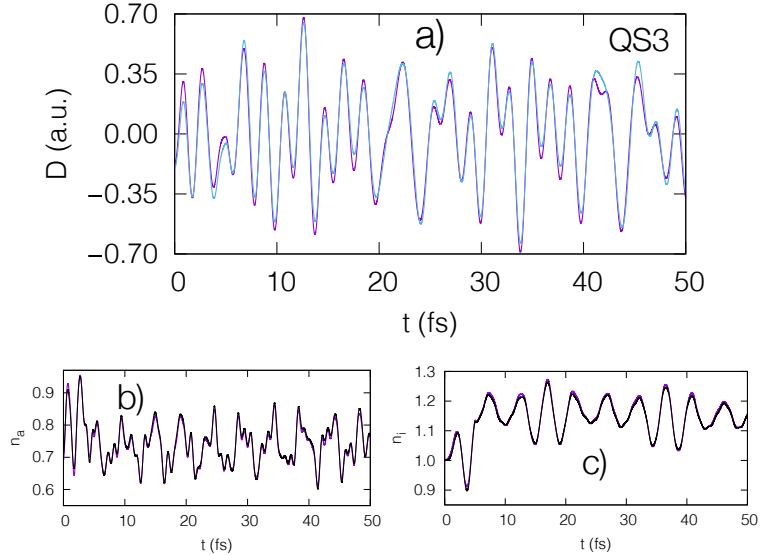}
\caption{Effect of regularization on dipole and population evolution, starting from QS3. a), TD
dipole where the purple and blue lines refer to the numerically exact and the {\slshape regularized}
SR theory results, respectively. b) and c) show the comparison between the regularized (black line)
and unitary (purple line) reference for levels $a$ and $i$, correspondingly.}\label{qs3}
\end{figure}

Our SR CC theory results provide support for equation (\ref{psiApsi_ii}), which describes the evolution of a
quantum mechanical observable. To further test equation (\ref{psiApsi_ii}), we now examine the response of
states that derive from the regularized ground state and the diagonalization of $\tilde{\mc{A}}$.
Our regularization technique approximates energies and CC vectors, and thereby the associated
TD observables. However, it can provide useful results to analyze and understand quantum mechanical
quantities. We examine here the case where $\alpha$ is relatively large (compared to
$\Delta\epsilon$), $\alpha=4.0~\rm{eV}$. For
the iterative calculation of the $t$-amplitudes, in the quasi-Newton step, which involves energy
differences in a series of denominators, the regularization number broadens such energy differences
by its value, for example, $\epsilon_a-\epsilon_i+\alpha$, and any other energy difference between
the configurations. So if the gap between virtual and occupied levels is relatively small, the regularization
removes potential instabilities and yields convergent, approximate $t,~\Lambda$ amplitudes. Tables S1
and S2 (supplemental material) show the regularized and exact energy values for the system
considered in figure \ref{scheme}. We note that with greater regularization values, the errors
increase slightly. The errors, have an upper bound because the larger the $\alpha$ number is, the
more negligible the resulting $\hat{T}$ and $\hat{\Lambda}$ vectors become. Nonetheless, there are
elements in our theory to improve (not explored in this work) the accuracy of a regularized method.
Regularization, in our opinion, is convenient for cases where instabilities arise in the
determination of the ground-state cluster amplitudes. To study the response of the system after the
ground-state regularized amplitudes are found, we follow the same steps as for the unregularized
propagations discussed before. Figure \ref{qs3} shows the evolution of the observable $\hat{D}$ for
the regularized system with $\alpha=4.0~\rm{eV}$. The system in this case begins from the state
QS3. As opposed to the unregularized calculations, small deviations are slightly more noticeable in
this case due to the effect of the regularization. This can be observed for the dipole, figure
\ref{qs3}.a, and the populations of the $a$ and $i$ levels, figures \ref{qs3}.b and .c,
respectively. Such deviations are caused by fluctuations in the excitation energies with respect to
the exact ones, and in the EOM-CC wavefunctions. 


As mentioned in Ref. \cite{mosquera2022excited}, the present SR CC approach is asymmetric
and, {\slshape in its current form}, is unable to reproduce {\slshape exactly} the standard
wavefunction theory results, despite these agree strikingly well. Nonetheless, the robustness of the
non-Hermitian SR CC model holds even if the strength of the electron-electron interaction is
increased, as shown in figure S1 (supplemental material). Repeating the steps outlined in
section \ref{comput}, but now for $b=0.25~\rm{eV}$ and $w_0=U_0=0.5~\rm{eV}$, we observe the same kind of overall
agreement between the SR theoretical propagation and the unitary reference evolution. The
ground-state amplitudes are in this scenario in the range $[0.08-0.2]$, reflecting a strong
electron-electron interaction (Tables S3 and S4, supplemental material). Similarly, the
regularization method yields the TD quantum dipole in close agreement with both the unregularized SR
and standard quantum mechanical results. Deviations with respect to standard simulations occur in
our theory because the SR CC wavefunctions display a type of normalization that does not take place
in the parent quantum mechanical theory. Even though one can normalize wavefunctions in a form that
complies with the Hermitian theory, this would require numerical operations that work against the
practical purpose of asymmetric CC theories. However, the unitary CC method is an alternative to the
non-Hermitian CC theory that is widely applied in quantum algorithms for simulations in chemistry.
The SR theory is applicable to expand TD unitary CC theory, but it necessitates a separate careful
analysis not covered in this work.  Nonetheless, the present SR theory captures accurately the TD
behavior of the observable represented by $\hat{A}$.

Returning to potential applications of the SR CC theory in realistic cases, for strong external
perturbations that are turned on for long enough such that they induce a significant response by the
quantum system, high-order excitation vectors of fourth order or beyond could be necessary. In these
situations, an alternative approach to avoid such high-order excitation vectors is the use of TD
relaxed orbitals \cite{kvaal2012ab,sato2018communication,pathak2021time}.  For example, a
possibility to expand the SR formalism in this direction is to employ auxiliary reference TD
orbitals that derive from a TD variational method. Such orbitals could also be treated as functions
of the parameters $g_{\rm{L}}$ and $g_{\rm{R}}$.  These would in turn provide a basis to find a
different set of TD equations for the excitation vectors that may be truncated at the third or
second order.

Our proposed SR CC theory derives from the standard (non-Hermitian yet size-extensive) TD CC
formalism \cite{koch1990coupled}, so it can propagate initial states described by linear quantum superpositions. In
practical TD propagations, SR CC methods might inherit the numerical challenging technicalities of
standard TD CC, which arise from its non-linear dependencies on the standard TD excitation vectors.
TD EOM-CC theory, in contrast, offers a more direct approach to propagations that does not
rely on the non-linear aspects of standard TD CC methods, but it is not size-extensive; TD EOM-CC
operates similarly as TD configuration-interaction theory \cite{sonk2011td,nascimento2016linear}. On the other
hand, the non-Hermiticity of standard TD CC \cite{pedersen1997coupled} has the
undesirable issue of breaking strict time-reversibility, which is a fundamental physical constraint,
but, as mentioned before, it is a matter that can be resolved in unitary CC theories, or lessened
with an external symmetrization \cite{pedersen1998time}. Finally, the starting (initial-state)
transition elements in our theory requires, in principle, the application of conventional response
theory, which can be computationally demanding for large systems, but still remains size extensive.
For simplicity we expressed these elements in a sum-over-states fashion, which requires EOM-CC
eigenvectors and eigenvalues, but evidences the fact that these can be computationally expensive.
However, approximations or approaches that circumvent the need for response-theory relaxation terms
for initial observable matrices, such as dipole matrices, could improve computational efficiency and
be the subject of future work.

\section{Conclusion}
This work examines in depth second response theory, a time-dependent coupled-cluster formalism to
propagate general quantum superpositions, and presents a simple formula, not developed previously,
that predicts accurately the evolution of a quantum mechanical observable of interest, including
probabilities and coherences. Our theory gives connected expressions to determine observables in the
non-linear regime, and requires the solution to the standard ground-state propagation problem and
the time-dependency of three response excitation vectors. With these elements provided, we studied
its application to a model three-electron/three-level system under different initial quantum
superpositions. This type of propagation is relevant to further developing theoretical approaches
and the study of systems that are thermally and/or optically excited. Finally, we examined the use of
the SR energy expressions to approach regularized systems, which do not solve exactly the
ground-state problem, but offer stable approximations to its cluster amplitudes, excited-state
energies, and CC excitation vectors. These could lead to further improvements to cases where
determining the cluster amplitudes is difficult because of low energy gaps, or related adverse
effects. 

\ack M.A.M. thanks Montana State University, Bozeman, for startup funding. The author also acknowledges
support by the National Science Foundation through the MonArk Quantum Foundry, DMR-1906383. 

\section*{References}
\bibliographystyle{iopart-num}
\bibliography{refs}

\end{document}


\title[Supplemental material]{Supplemental material. Second response theory: A theoretical formalism
for the propagation of quantum superpositions}

\author{Mart\'in A. Mosquera}
\address{Department of Chemistry and Biochemistry, Montana State University, Bozeman, MT 59718,
USA}
\ead{martinmosquera@montana.edu}

\maketitle

\begin{figure}[htb!]
\centering
\includegraphics[scale=0.7]{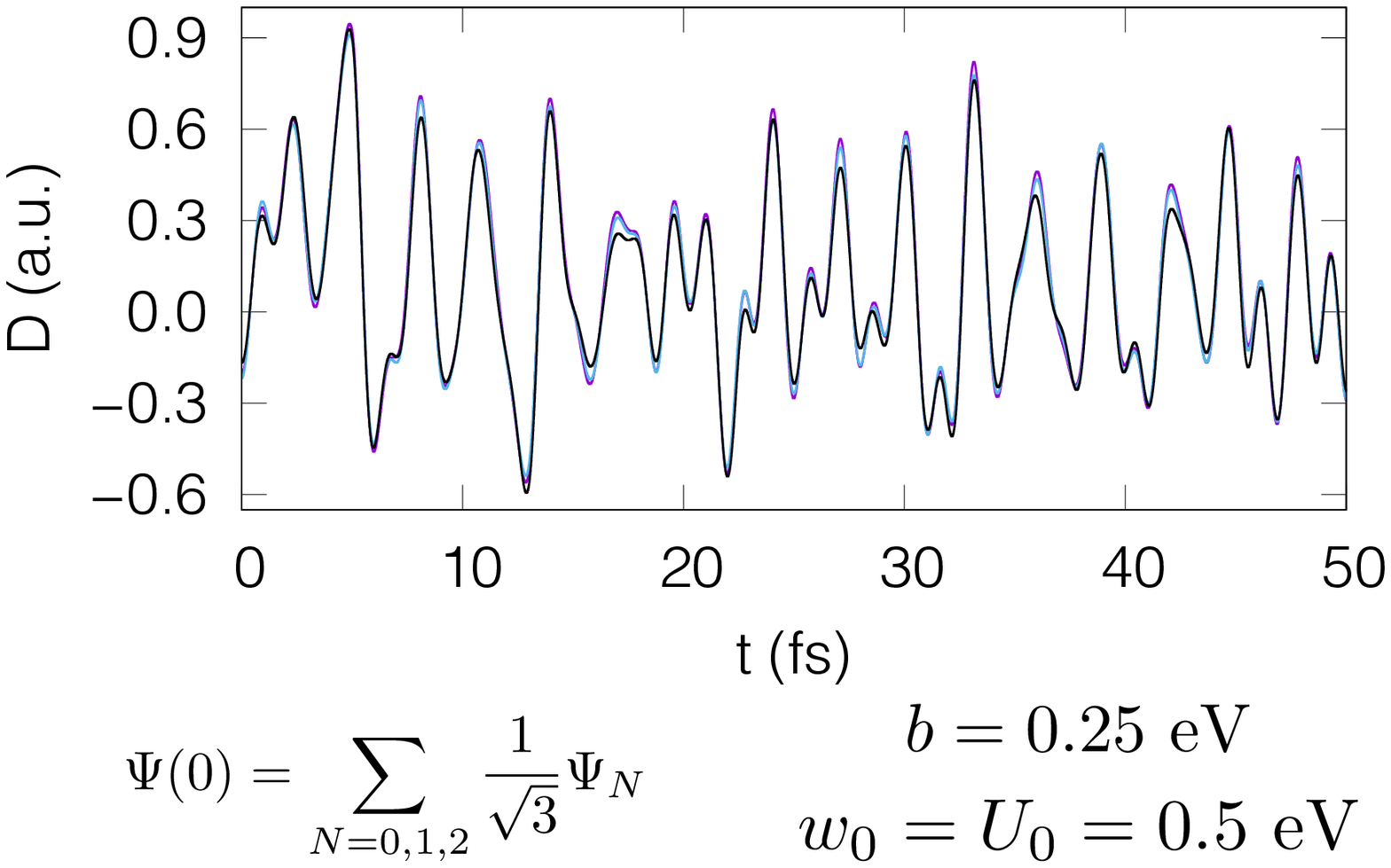}
\caption{Evolution remains robust upon increasing the electron-electron interaction strength. Purple
line: exact result, blue line: SR theory (no regularization), black line: regularized SR results
($\alpha=4.0~\rm{eV}$).}\label{qs1_uee}
\end{figure}

\section{Extra equations}
The TD cluster operators $\hat{\lambda}^{\rm{E}}_{\rm{r}}$, $\hat{\lambda}^{\rm{E}}_{\rm{l}}$, and
$\hat{\lambda}^{\rm{E}}_{\rm{l,r}}$ follow the relations:
\begin{eqnarray}\label{all_eqs}
-\ui\partial_t \lambda_{\rm{l},\mu}^{\rm{E}}(t)&=
\langle \hat{\lambda}_{\rm{l}}^{\rm{E}}(t)
[\bar{H}_x(t),\hat{\tau}_{\mu}]\rangle_0~,\\
-\ui\partial_t \lambda_{\rm{r},\mu}^{\rm{E}}(t)&=
\langle \hat{\lambda}(t)[\bar{H}_{x,\tau,\mu}(t),
\hat{x}_{\rm{r}}(t)]
+\hat{\lambda}_{\rm{r}}^{\rm{E}}(t)\bar{H}_{x,\tau,\mu}(t)\rangle_0~,\\
-\ui\partial_t \lambda_{\rm{l,r},\mu}^{\rm{E}}(t)&=\langle \hat{\lambda}_{\rm{l,r}}^{\rm{E}}(t)
\bar{H}_{x,\tau,\mu}(t)
+\hat{\lambda}_{\rm{l}}^{\rm{E}}(t)
[\bar{H}_{x,\tau,\mu}(t),\hat{x}_{\rm{r}}(t)]
\rangle_0~.
\end{eqnarray}

\newpage

\section{Tables}
\begin{table}[htbp]
  \centering
  \caption{Energies the eigen-states of the unperturbed 3-electron/3-level system for the case where $b$, $w_0$
  and $U_0$ are 0.1, 0.2, and 0.2 eV, respectively. For each state we show the most dominant
  configurations (Figure 1). We denote the dominant configurations simply as ``Config.'' below. For
  example, wavefunction (WF) $\Psi_3$ is dominated by the 6th and 2nd configurations. We also show
  the effect of regularization for three different values of $\alpha$, 0.5, 4.0, and 8.0 eV.}
  \vspace{5pt}
    \begin{tabular}{c|c|c|c|c|c}
    WF No. &  Energy (a.u.)  & Config. & $E_N$, $\alpha=0.5~\mathrm{eV}$ & $E_N$,
    $\alpha=4.0~\mathrm{eV}$  & $E_N$, $\alpha=8.0~\mathrm{eV}$  \\
    \hline
    0     & 0.03406112 & 0     & 0.03450364 & 0.03567967 & 0.03607546 \\
    1     & 0.07313581 & 1     & 0.07321405 & 0.07338131 & 0.07342816 \\
    2     & 0.08035159 & 3     & 0.08047917 & 0.08070788 & 0.08076523 \\
    3     & 0.10975262 & 6, 2  & 0.10986812 & 0.11009739 & 0.11015831 \\
    4     & 0.10994818 & 2, 6  & 0.11002712 & 0.11016849 & 0.11020221 \\
    5     & 0.11190407 & 4     & 0.11153361 & 0.11076561 & 0.11055775 \\
    6     & 0.15482654 & 7, 5  & 0.15471626 & 0.15449552 & 0.15443684 \\
    7     & 0.15583540 & 5, 7  & 0.15558061 & 0.15493385 & 0.15471786 \\
    8     & 0.19183032 & 8     & 0.19172308 & 0.19141593 & 0.19130383 \\
    \end{tabular}%
\end{table}%

\begin{table}[htbp]
  \centering
  \caption{Ground-state cluster amplitudes ($\{t_{\mu}\}$) of the 3-electron/3-level system for the
  case where $b$, $w_0$ and $U_0$ are 0.1, 0.2, and 0.2 eV, respectively. We also show the effect of
  regularization on these amplitudes for three different values of $\alpha$, 0.5, 4.0, and 8.0 eV.}
  \vspace{5pt}
    \begin{tabular}{c|c|c|c|c}
    Configuration  & ~~$t_{\mu}$ (unitless)~~    & $t_{\mu}'$, $\alpha=0.5~\mathrm{eV}$ & $t_{\mu}'$, $\alpha=4.0~\mathrm{eV}$  & $t_{\mu}'$, $\alpha=8.0~\mathrm{eV}$  \\
    \hline
    1     & -0.07950747 & -0.05601643 & -0.01868837 & -0.01067200 \\
    2     & -0.04329081 & -0.03546118 & -0.01573367 & -0.00964238 \\
    3     & -0.06698762 & -0.04964707 & -0.01797352 & -0.01044045 \\
    4     & -0.09473071 & -0.07728911 & -0.03312300 & -0.01995302 \\
    5     & -0.06063376 & -0.05290738 & -0.02763888 & -0.01782138 \\
    6     & -0.04330929 & -0.03546662 & -0.01573373 & -0.00964238 \\
    7     & -0.06079416 & -0.05299412 & -0.02764646 & -0.01782314 \\
    8     & -0.04665203 & -0.04190695 & -0.02428942 & -0.01636484 \\
    \end{tabular}%
\end{table}%

\begin{table}[htbp]
  \centering
  \caption{Energies of the eigen-states of the unperturbed 3-electron/3-level system for the case
  where $b$, $w_0$ and $U_0$ are 0.25, 0.5, and 0.5 eV, respectively. For each state we show the
  most dominant configurations (Figure 1).  We denote the dominant configurations simply as
  ``Config.'' below.  For example, $\Psi_7$ is dominated by the 5th and 7th configurations. We also
  show the effect of regularization for three different values of $\alpha$, 0.5, 4.0, and 8.0 eV.}
  \vspace{5pt}
    \begin{tabular}{c|c|c|c|c|c}
    WF No. &  Energy (a.u.)  & Config. & $E_N$, $\alpha=0.5~\mathrm{eV}$ & $E_N$,
    $\alpha=4.0~\mathrm{eV}$  & $E_N$, $\alpha=8.0~\mathrm{eV}$  \\
    \hline
    0     & 0.0236728 & 0     & 0.02523072 & 0.03055085 & 0.03271333 \\
    1     & 0.0704724 & 1     & 0.07096318 & 0.07230300 & 0.07275556 \\
    2     & 0.0877195 & 3     & 0.08833656 & 0.09007605 & 0.09071326 \\
    3     & 0.1074657 & 6, 2  & 0.10792629 & 0.10915896 & 0.10957066 \\
    4     & 0.1088416 & 2, 6  & 0.10907914 & 0.10971073 & 0.10991757 \\
    5     & 0.1193940 & 4     & 0.11809707 & 0.11421255 & 0.11277842 \\
    6     & 0.1678397 & 7, 5  & 0.16744433 & 0.16635852 & 0.16598905 \\
    7     & 0.1733486 & 5, 7  & 0.17234761 & 0.16909455 & 0.16779473 \\
    8     & 0.2069911 & 8     & 0.20632056 & 0.20428027 & 0.20351289 \\
    \end{tabular}%
\end{table}%

\begin{table}[htbp]
  \centering
  \caption{Ground-state cluster amplitudes ($\{t_{\mu}\}$) of the 3-electron/3-level system for the
  case where $b$, $w_0$ and $U_0$ are 0.25, 0.5, and 0.5 eV, respectively. We also show the effect
  of regularization on these amplitudes for three different values of $\alpha$, 0.5, 4.0, and 8.0
  eV.}
  \vspace{5pt}
    \begin{tabular}{c|c|c|c|c}
    Configuration  & ~~$t_{\mu}$ (unitless)~~    & $t_{\mu}'$, $\alpha=0.5~\mathrm{eV}$ & $t_{\mu}'$, $\alpha=4.0~\mathrm{eV}$  & $t_{\mu}'$, $\alpha=8.0~\mathrm{eV}$  \\
    \hline
    1     & -0.12821361 & -0.10064334 & -0.04148473 & -0.02494136 \\
    2     & -0.08356321 & -0.07146805 & -0.03567552 & -0.02271260 \\
    3     & -0.09336382 & -0.07877364 & -0.03782335 & -0.02364346 \\
    4     & -0.20069546 & -0.17111214 & -0.08042701 & -0.04931169 \\
    5     & -0.12600387 & -0.11349122 & -0.06485739 & -0.04297576 \\
    6     & -0.08380465 & -0.07156498 & -0.03567760 & -0.02271279 \\
    7     & -0.12667178 & -0.11397311 & -0.06493685 & -0.04299760 \\
    8     & -0.10135496 & -0.09296701 & -0.05748125 & -0.03959413 \\
    \end{tabular}%
\end{table}%